\documentclass[aps,prd,reprint,superscriptaddress,nofootinbib]{revtex4-2}
\usepackage{mathtools}
\usepackage{lineno}
\usepackage{graphicx} 
\graphicspath{{./images/}}
\usepackage[boxsize=0.8em,centertableaux]{ytableau}
\usepackage{amsmath, amssymb, amsthm}
\usepackage{braket}
\usepackage{tikz} 
\numberwithin{equation}{section}  
\usepackage{mdframed}
\usepackage{xcolor} 
\usepackage{amsfonts}
\usepackage{empheq}
\usepackage{mathrsfs}
\usepackage{mathtools}

\usepackage[colorlinks=true, allcolors=blue]{hyperref}
\newtheorem{Lemma}{Lemma}

\usepackage[bottom]{footmisc} 
\begin{document}

\title{Charge functions for all dimensional partitions}

\author{Hao Feng}
\email{fenghaozi@shu.edu.cn}
\affiliation{Department of Physics and Institute for Quantum Science and Technology, Shanghai University, Shanghai 200444, China}

\author{Tian-Shun Chen}
\email{cts2003912@shu.edu.cn}
\affiliation{Department of Physics and Institute for Quantum Science and Technology, Shanghai University, Shanghai 200444, China}

\author{Kilar Zhang}
\email{kilar@shu.edu.cn; Corresponding Author}
\affiliation{Department of Physics and Institute for Quantum Science and Technology, Shanghai University, Shanghai 200444, China}
\affiliation{Shanghai Key Lab for Astrophysics, Shanghai 200234, China}
\affiliation{Shanghai Key Laboratory of High Temperature Superconductors, Shanghai 200444, China}

\begin{abstract}
The charge functions for $n$-dimensional partitions are known for $n=2,3,4$ in the literature. In a recent work, we gave the expression for arbitrary odd dimension; here we further conjecture a formula for all even-dimensional cases. This conjecture is proved rigorously for 6D, and numerically verified for 8D.
\end{abstract}

\maketitle


\section{Introduction}
The statistical enumeration of Bogomol'nyi-Prasad-Sommerfield (BPS) states on toric Calabi-Yau $n$-folds stands as a fundamental problem in both topological string theory and enumerative geometry~\cite{ooguri2009crystal, okounkov2006quantum, iqbal2008quantum}. The key to this program is the identification of $n$-dimensional partitions~\cite{macmahon1898memoir, macmahon1912ix}, which provide a combinatorial realization of the BPS Hilbert space. In these systems, the molten crystal growth is governed by the underlying BPS algebra~\cite{harvey1996algebras, harvey1998algebras, bourgine2019note}, a symmetry structure whose representation theory is encoded within the spectral properties of its Cartan operators. The eigenvalues of these operators are packaged into a single meromorphic object: the charge function $\psi(u)$. As a decisive factor in the algebraic bootstrap procedure, $\psi(u)$ determines the melting rules of the crystal, where its simple poles are in one-to-one correspondence with the box positions where a box can be added to or removed from a partition~\cite{li2020quiver, galakhov2024charging}.

From a mathematical viewpoint, the relevance of such combinatorial objects is
not restricted to the usual low-dimensional instanton-counting examples.
Partitions already appear naturally in number theory and representation
theory, and Young diagrams are closely related to Hilbert schemes of points on
surfaces. Plane partitions provide the local combinatorics behind the
Donaldson--Thomas theory of toric Calabi--Yau threefolds. In dimension four,
solid partitions similarly enter the study of local Calabi--Yau fourfold
geometry and zero-dimensional Donaldson--Thomas invariants. We therefore view
higher-dimensional charge functions as potential local combinatorial data for
exploring analogous enumerative and representation-theoretic structures on
toric Calabi--Yau $n$-folds, while keeping in mind that the corresponding
physical interpretation above dimension four is less established
~\cite{andrews1998theory,nakajima1999lectures,okounkov2007random,donaldson1998gauge,kontsevich2011cohomological,maulik2006gromov,nekrasov2019magnificent,nekrasov2020magnificent,cao2018zero}.

While the charge functions for low-dimensional partitions are well-established, they exhibit an evolution toward increasing non-locality. For 3D partitions, the affine Yangian of $\mathfrak{gl}_1$ successfully describes the dynamics through simple factorized rational kernels~\cite{prochazka2016,rapcak2020cohomological,galakhov2024charging}. This framework precisely describes the algebraic structure behind Nekrasov instanton partition functions and the AGT correspondence~\cite{nekrasov2003seiberg, alday2010liouville}. However, moving toward 4D solid partitions and higher dimensions necessitates the inclusion of collective excitations known as box clusters~\cite{nekrasov2019magnificent, nekrasov2020magnificent, galakhov2024charging}. This transition reflects the growing complexity of the instanton moduli space and the associated equivariant cohomology~\cite{schiffmann2013cherednik, arbesfeld2013presentation, nekrasov2003seiberg, donaldson1998gauge, kontsevich2011cohomological}.
Building upon these developments, in our preceding work \cite{xiang2025chargefunctionshighdimensional}, a general formula is proposed for arbitrary odd-dimensional partitions, which is rigorously proved for the 5D case, and tested for the 7D and 9D cases.

However, generalizing these results to arbitrary even dimensions ($n=2K, K \geq 3$) remains a formidable task. The primary difficulty lies in the asymmetric distribution induced by odd-order product terms and the breaking of pole-sign symmetry in even dimensions, which prevents the charge function from following the simple patterns observed in lower dimensions~\cite{macmahon2004combinatory, xiang2025chargefunctionshighdimensional}. In this letter, we overcome these obstacles by proposing a universal conjecture for the charge functions of even-dimensional partitions. We provide a rigorous proof for the 6D case by introducing a lemma and applying an exhaustive enumeration method~\cite{bonelli2023adhm, szabo2023instanton}, and perform extensive numerical verification via Monte Carlo sampling for 8D partitions. Compared with the proof of the 4D charge function in Ref.~\cite{galakhov2024charging}, which checks local projected configurations explicitly, our proof follows the induction strategy of Ref.~\cite{xiang2025chargefunctionshighdimensional} and reduces the problem to a background partition plus a local hypercube contribution. Our formula completes the library of charge functions for all dimensions, providing the essential data required to explore the BPS algebras of higher-dimensional Calabi-Yau manifolds~\cite{cao2018zero, maulik2019quantum}.

\section{partitions}
First let us collect the necessary definitions relating partitions, following the descriptions in \cite{xiang2025chargefunctionshighdimensional, galakhov2024charging}.

An $n$-dimensional partition $\Delta^{(n)}$ should  obey the melting rule:
For any box $\vec{\square} \in \mathbb{Z}^n_{\geq 0}$ and any $k = 1, 2,\cdots, n$, if there exists $\vec{\square}' \in \Delta^{(n)}$ with $\vec{\square}' = \vec{\square} + \vec{e}_k$, then $\vec{\square} \in \Delta^{(n)}$.

Equivalently, an $n$-dimensional partition is a finite subset of the non-negative
lattice that is closed toward the origin: whenever a box is present, all boxes
obtained from it by moving one step in a negative coordinate direction must
also be present. Thus ordinary Young diagrams, plane partitions, and solid
partitions are respectively the cases $n=2,3,4$ of the same definition.

\begin{figure}
    \centering
    \includegraphics[width=1\linewidth]{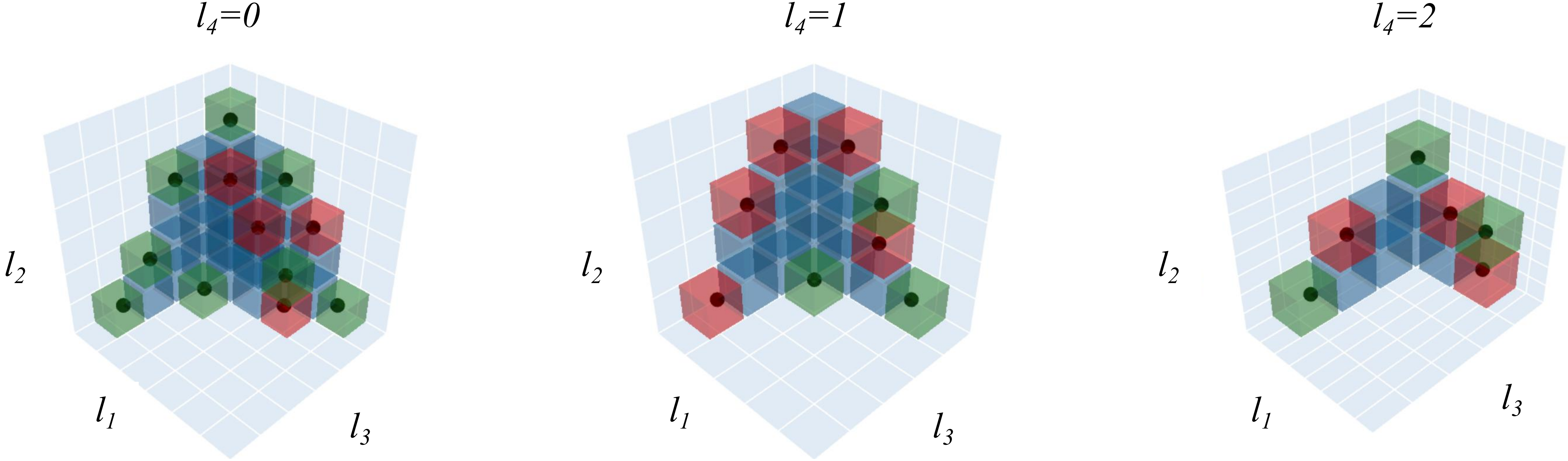}
    \caption{Schematic of a 4D partition. Green and red designate addable and removable box positions, respectively, while black dots indicate sites where our method predicts a box (either red or green) to be. The subfigures show slices at $l_4 = 0, 1, 2$ from left to right, all of which are consistent with the melting rule.}
    \label{fig:partition}
\end{figure}

For instance,  Fig.\ref{fig:partition} presents a schematic of an $n=4$ partition satisfying the melting rule.

The canonical basis is defined as
\begin{equation}
\vec{e}_i = \left( \delta_{1i}, \delta_{2i}, \dots, \delta_{ni} \right), \quad i = 1, 2, \dots, n,
\end{equation}
with $\delta_{ij}$ the Kronecker delta. When all $n$ coordinate
directions are involved, we write the diagonal lattice vector as
\begin{equation}
    \vec q_n\coloneqq\sum_{i=1}^{n}\vec e_i .
\end{equation}

A set of complex numbers $h_i$, called weights or flavor parameters, characterizes the equivariant toric action on CY$_n$ as $(x_1,x_2,\cdots,x_n)\mapsto (e^{h_1}x_1,e^{h_2}x_2,\cdots,e^{h_n}x_n)$. They satisfy the Calabi-Yau condition:
$
    \sum_{i=1}^{n}h_i=0.
$

This condition is part of the toric Calabi--Yau equivariant setting used
throughout this paper. The same convention is used in the lower-dimensional
charge-function formulas; for example, Ref.~\cite{galakhov2024charging} gives
the Young-diagram case with $h_1+h_2=0$ and the plane-partition case with
$h_1+h_2+h_3=0$. In this setting the projected coordinate of a box is the
corresponding flavor charge in the equivariant weight space.

Here two fundamental sets are of particular importance:
\begin{equation}
\operatorname{A}_{\Delta^{(n)}} \subset \mathbb{Z}^n_{\geq 0} \quad \text{and} \quad \operatorname{R}_{\Delta^{(n)}} \subset \mathbb{Z}^n_{\geq 0},
\end{equation}
which consist of all boxes in the lattice $\mathbb{Z}^n_{\geq 0}$ that can be added to or removed from $\Delta^{(n)}$, respectively, while ensuring the resulting configuration remains a valid $n$-dimensional partition.

For example, for the one-box partition $\Delta^{(n)}=\{\vec 0\}$, the only
removable box is $\vec 0$, while the addable boxes are
$\vec e_1,\ldots,\vec e_n$.

We now define a projection map $c$ as follows\footnote{Throughout this work, the coordinate components of a box $\vec{\square}$ are denoted by $l_i$.}:
\begin{equation}
\label{eq:c=le}
    c:\quad \vec{\square}=\sum_{i=1}^{n}l_i \vec{e}_i\longmapsto \sum_{i=1}^{n}l_ih_i,
\end{equation}
where the weight parameters $\{h_i\}_{i=1,\dots,n}$ are generic complex numbers satisfying the Calabi-Yau condition.

We denote the image of the projection map by $\mathcal P$. Throughout this
work, $c$ denotes the map itself, while $x,x'\in\mathcal P$ denote points in
the projected space.
For a box $\vec{\square}=\sum_{j=1}^{n}l_j\vec e_j$, we write
$l_i(\vec{\square})=l_i$, where $l_i\in\mathbb Z_{\geq0}$. Because of the
Calabi--Yau condition,
\begin{equation}
    c(\vec{\square}+\vec q_n)=c(\vec{\square}),
\end{equation}
so the projection identifies lattice points that differ by a diagonal shift.
Hence the coordinate representation of a projected point is not unique. We
choose a unique normalized representative by defining
\begin{equation}
    l'_i=l_i-\min_{1\leq j\leq n}l_j.
\end{equation}
The projected point can then be written uniquely as
\begin{equation}
 x = \sum_{i=1}^{n} l'_i h_i,
 \qquad l'_i \geq 0,
 \qquad \min_i l'_i=0.
\end{equation}

We denote the normalized coordinates of $x$ by $l'_i(x)=l'_i$.
For instance, in the case $n=3$, the two boxes with coordinates $(2,1,0)$ and
$(3,2,1)$ differ by $\vec q_3$ and therefore have the same projected
coordinate. The normalized representative is obtained by subtracting the
minimal coordinate from all components.

We define $P_n$ as the set of all possible $n$-dimensional partitions and $\mathcal{P}$ as the set of points in the projected space:
\label{sec:conjecture chare function}
\begin{equation}
    P_n\coloneqq\left \{ \Delta^{(n)} \ \text{for any number of boxes} \right \} \subset 2^{\mathbb{Z}_{\geq 0}^{n} },
\end{equation}
\begin{equation}
    \mathcal{P} \coloneqq \left\{ x = \sum_{i=1}^{n}l_ih_i
    \;\middle|\; l_i \in \mathbb{Z}_{\geq0} \right\}
    \cong \mathbb{Z}^{n-1}.
\end{equation}

It is useful to collect the boundary positions of a partition into
\begin{equation}
    C_{\Delta^{(n)}}\coloneqq
    \operatorname{A}_{\Delta^{(n)}}\cup
    \operatorname{R}_{\Delta^{(n)}},
\end{equation}
and to denote their image in the projected space by
\begin{equation}
    D_{\Delta^{(n)}}\coloneqq
    \left\{c(\vec{\square})\in\mathcal P
    \;\middle|\;
    \vec{\square}\in C_{\Delta^{(n)}}\right\}.
\end{equation}

Finally, we introduce a set of certain partitions at a given box $\vec{\square}$, denoted by $G(\vec{\square})$. This set comprises all $n$-dimensional partitions that possess at least one addable or removable position whose projection coincides with $c(\vec{\square})$:
\begin{equation}
  G(\vec{\square}) = \{\Delta^{(n)}\in P_n \mid \exists \tilde{\vec{\square}}\in \mathrm{A}_{\Delta^{(n)}} \cup \mathrm{R}_{\Delta^{(n)}}, c(\tilde{\vec{\square}})=c(\vec{\square})\}.
\end{equation}

Here $c(\tilde{\vec{\square}})$ is the projected coordinate of
$\tilde{\vec{\square}}$ under the same projection map $c$ defined in
Eq.~\eqref{eq:c=le}.

\section{charge functions}
In this letter, we show that the charge functions $\psi(u)$ for arbitrary dimensional partitions $\Delta^{(n)}$ can be explicitly constructed. These functions are characterized by the following fundamental properties:

\begin{mdframed}
\begin{enumerate}
    \item \label{charge function property 1}
    $\psi_{\Delta^{(n)}}(u)$ is a meromorphic function of $u$ whose singularities are restricted to simple poles.
    \item \label{charge function property 2}
    There exists a one-to-one correspondence between the poles of $\psi_{\Delta^{(n)}}(u)$ and the projected coordinates $c(\vec{\square})$ associated with the addable and removable boxes $\vec{\square} \in \mathrm{A}_{\Delta^{(n)}} \cup \mathrm{R}_{\Delta^{(n)}}$.
\end{enumerate}
\end{mdframed}

The two properties above do not determine the charge function uniquely. They
specify only the required pole structure: for a given partition, the projected
addable/removable surface positions must appear as simple poles, and no other
projected positions should give poles. In this sense, one could always define a
minimal rational function whose only poles are simple poles at those projected
surface positions, for example by taking a product over them. Such a
construction would satisfy the pole criterion, but it would be a global object
depending directly on the full surface structure of the partition.

The point of the present work is more restrictive. We look for charge functions
which are built multiplicatively from local contributions of boxes and finite
box clusters. This locality is essential from the viewpoint of the BPS-algebra
bootstrap, because the charge function is expected to arise as the eigenvalue
of a Cartan current whose action factorizes when boxes or clusters are added.
Uniqueness is not assumed or required in our construction. Even within the
class of factorized expressions, we do not claim a uniqueness theorem. Our goal
is to exhibit one explicit local product formula that reproduces the
melting-rule pole structure and is therefore suitable for the algebraic
construction.

The lower-dimensional charge functions have already been established in the
literature. In Appendix~\ref{appendix:lower_dim} we recall the cases
$n=2,3,4$ in the notation used here. In particular, the two-dimensional
Young-diagram formula, the three-dimensional plane-partition formula, and the
four-dimensional solid-partition formula are given in
Ref.~\cite{galakhov2024charging}. The three-dimensional case has the standard
interpretation as the eigenvalue of the Cartan current in the affine Yangian of
$\mathfrak{gl}_1$, while the four-dimensional case requires additional cluster
contributions. Our higher-dimensional formulas are proposed as extensions of
this established Calabi--Yau projected-weight-space framework.

In the following, we directly present the general expressions of the charge functions for arbitrary dimension $n$, distinguishing between the even and odd cases. A review of the charge functions for lower-dimensional partitions can be found in Appendix \ref{appendix:lower_dim}. Some elementary examples for $n=2,3,4$ are collected in Appendix~\ref{appendix:low_dim_examples}.
\vspace{2em}
\paragraph{Formula for arbitrary odd dimension}(proposed in \cite{xiang2025chargefunctionshighdimensional}, proved for 5D, and checked for 7D and 9D):
In dimension $n=2K+1,K\geq1$, the charge function
reads
\begin{equation}
    \psi_{\Delta^{(n)}}(u)=\psi_{0}(u)\psi'_{\Delta^{(n)}}(u),
\end{equation}
with
$
    \psi_{0}=\frac{1}{u},
$ and
\begin{equation}
\begin{split}
    &\psi'_{\Delta^{(n)}}(u)=\prod_{\vec{\square}\in \Delta^{(n)}}\varphi^{(2K+1)}_1(u-c(\vec{\square}))\\
    &\times\prod_{m=2}^{K}\prod_{\phi_{2m}\subset \Delta^{(n)}}\varphi^{(2K+1)}_{2m}(u-c(\phi_{2m})),
\end{split}
\end{equation}
where
\begin{equation}
\varphi^{(2K+1)}_1(u)=\frac{\prod_{m=1}^{K}\prod_{1\leq a_1<a_2<\cdots<a_{2m}\leq 2K+1}(u-\sum_{i=1}^{2m}h_{a_{i}})}{\prod_{i=1}^{2K+1}(u-h_i)},
\end{equation}
\begin{equation}
    \varphi^{(2K+1)}_{2m}(u)=\frac{1}{u}.
\end{equation}

In the above, for a base box $\vec{\square}$ and an index set
$S=\{s_1,\ldots,s_{p-1}\}\subseteq\{1,\ldots,n\}$ with
$1\leq s_1<\cdots<s_{p-1}\leq n$, we define the $p$-box cluster
\[
\phi_p(\vec{\square};S)
\coloneqq
\left\{\vec{\square},
\vec{\square}+\vec{e}_{s_1},\ldots,
\vec{\square}+\vec{e}_{s_{p-1}}\right\}.
\]
The condition $\phi_p(\vec{\square};S)\subseteq\Delta^{(n)}$ means that both
the base box and all $p-1$ shifted boxes belong to the partition. Accordingly,
the shorthand product $\prod_{\phi_p\subseteq\Delta^{(n)}}$ runs over all base
boxes $\vec{\square}$ and all allowed index sets $S$ for which this condition
holds. We associate to such a cluster the projected coordinate
\begin{equation}
    c\bigl(\phi_p(\vec{\square};S)\bigr)
    \coloneqq c(\vec{\square})+\sum_{i=1}^{p-1}h_{s_i}.
\end{equation}

This coordinate is the projection of the diagonally shifted lattice point
$\vec{\square}+\sum_{i=1}^{p-1}\vec e_{s_i}$, which is the spectral position
at which the corresponding cluster factor contributes.

\paragraph{Formula for arbitrary even dimension:}
We conjecture the form of the charge function in dimension $n=2K,K\geq2,$
\begin{equation}
    \psi_{\Delta^{(n)}}(u)=\psi_{0}(u)\psi'_{\Delta^{(n)}}(u),
    \label{charge function general}
\end{equation}
where again
$
    \psi_{0}=\frac{1}{u},
$
and
\begin{equation}
\begin{aligned}
    \psi'_{\Delta^{(n)}}(u) =
    & \prod_{\vec{\square}\in \Delta^{(n)}}\varphi^{(2K)}_1(u-c(\vec{\square})) \\
    & \times \prod_{m=2}^{K-1}\prod_{\phi_{2m}\subset \Delta^{(n)}}\varphi^{(2K)}_{2m}(u-c(\phi_{2m})) \\
    & \times \prod_{\phi_{2K}\subset \Delta^{(n)}}\varphi^{(2K)}_{2K}(u-c(\phi_{2K})) \\
    & \times \prod_{\phi_{2K+1}\subset \Delta^{(n)}}\varphi^{(2K)}_{2K+1}(u-c(\phi_{2K+1})).
\end{aligned}
\label{eq:psi'}
\end{equation}
and the exact forms for the factors are:
\begin{equation}
\begin{split}
   &\varphi^{(2K)}_1(u)=\frac{\prod_{i=1}^{2K}(u+h_i)}{\prod_{i=1}^{2K}(u-h_i)} \\
   \times\prod_{m=1}^{K-1}&\prod_{1\leq a_1<a_2<\cdots<a_{2m}\leq 2K}(u-\sum_{i=1}^{2m}h_{a_{i}}),
\end{split}
\end{equation}
\begin{equation}
    \varphi^{(2K)}_{2m}(u)=\frac{1}{u},(m\in\{2,3,...,K-1\}),
\end{equation}
\begin{equation}
    \varphi^{(2K)}_{2K}(u)=\frac{1}{u^2},
\end{equation}
and
\begin{equation}
    \varphi^{(2K)}_{2K+1}(u)=u^2.
\end{equation}

In the notation $\varphi_p^{(n)}$, the superscript specifies the dimension of
the partition and the subscript labels the size of the box cluster whose
contribution is represented by the factor. In particular, a factor $u^2$
contributes a zero of order two, whereas $u^{-2}$ contributes a pole of order
two, at the projected coordinate of the corresponding cluster.

It is easy to tell for $n=4$, i.e. $K=2$,  \eqref{charge function general} identifies with the known formula for solid partitions as shown in \eqref{4D charge function}.

It is worth mentioning that the 2D case \eqref{2D charge function} is an exception, since its dimension is too low to share the common structure of p-box clusters.

\section{Potential Function}
\label{sec:Potential Function}

To characterize the analytic structure of the charge functions, we introduce
a potential function $\omega_{0,\Delta^{(n)}}(x)$, which gives the order of
the pole at a projected point $x\in\mathcal P$:
\begin{equation}
    \omega_{0,\Delta^{(n)}}(x)=\begin{cases}
    m, & \text{for a pole of order } m, \\
    0, & \text{for an analytic point,} \\
    -m, & \text{for a zero of order } m.
\end{cases}
\end{equation}

The explicit form of $\omega_{0,\Delta^{(n)}}(x)$ derived from
\eqref{eq:psi'} consists of a vacuum contribution and a partition-dependent
term:
\begin{equation}
\omega_{0,\Delta^{(n)}}(x)=\delta_{0,x}+\omega_{\Delta^{(n)}}(x).
\end{equation}
where
\begin{equation}
\omega_{\Delta^{(n)}}(\tilde{x})
\coloneqq
\omega_{\Delta^{(n)},1}(\tilde{x})
+\omega_{\Delta^{(n)},\phi_{p}}(\tilde{x}).
\end{equation}
where the single-box contributions are given by:
\begin{equation}
\begin{aligned}
\omega_{\Delta^{(n)},1}(\tilde{x}) \coloneqq& \sum_{\vec{\square}\in \Delta^{(n)}} \biggl(
 \sum_{i=1}^n \delta_{\tilde{x}, c(\vec{\square})+h_i} - \sum_{i=1}^n \delta_{\tilde{x}, c(\vec{\square})-h_i} \\
& - \sum_{m=1}^{K-1}\,\sum_{1\leq a_1<\cdots<a_{2m}\leq n}
\delta_{\tilde{x},c(\vec{\square})+\sum_{r=1}^{2m}h_{a_r}} \biggr),
\end{aligned}
\end{equation}
and the cluster-dependent terms are defined via:
\begin{equation}
\label{def:potential of cluster}
\omega_{\phi_{p}}(\tilde{x}) \coloneqq \begin{cases}
    \delta_{\tilde{x},c(\phi_{p})} & p\in\{4,6,...,2K-2\} \\
    2\delta_{\tilde{x},c(\phi_{p})} & p=2K \\
    -2\delta_{\tilde{x},c(\phi_{p})} & p=2K+1
\end{cases},
\end{equation}
\begin{equation}
\omega_{\Delta^{(n)},\phi_{p}}(\tilde{x})
\coloneqq
\sum_{p\in\{4,6,...,2K\}\cup\{2K+1\}}
\,\sum_{\phi_{p}\subseteq \Delta^{(n)}}\omega_{\phi_{p}}(\tilde{x}).
\end{equation}
The potential function associated with a specific box $\vec{\square}$ is naturally defined as $\omega_{\Delta^{(n)}}(\vec{\square}) \coloneqq \omega_{\Delta^{(n)}}(c(\vec{\square}))$.

Using the potential function introduced above, we can rewrite the charge function properties as:
\begin{mdframed}
    \begin{enumerate}
        \item $\psi_{\Delta^{(n)}}(u)$ has only simple poles.
        \item
              $\Delta^{(n)} \in G(\vec{\square}) \, \,\Longleftrightarrow \, \,
                  \omega_{0,\Delta^{(n)}}(\vec{\square}) = 1.$
    \end{enumerate}
    \end{mdframed}

\textbf{Conjecture:}
The even dimensional charge function (\ref{charge function general}) satisfies the properties above.
\section{Proof}
\label{sec:Lemma}

The proof follows the reduction strategy used in our previous work
\cite{xiang2025chargefunctionshighdimensional}. Fix an arbitrary partition
$\Delta^{(n)}$ and a projected boundary position $c(\vec{\square})$ at which
we want to test the pole order. The reduction reviewed in
Appendix~\ref{appendix:hypercube_reduction} shows that, assuming
Lemma~\ref{Lemma5} below, the statement
$\omega_{0,\Delta^{(n)}}(\vec{\square})=1
\iff \Delta^{(n)}\in G(\vec{\square})$
can be compared with the same statement for a smaller partition obtained from
$\Delta^{(n)}$ by deleting a finite upper part. This deleted part is
determined by an elementary hypercube near $\vec{\square}$, together with the
additional boxes which must be removed so that the remaining set still obeys
the melting rule. The smaller partition is then handled by induction on
$|\Delta^{(n)}|$, while the difference between the original partition and the
smaller one is exactly the hypercube contribution controlled by
Lemma~\ref{Lemma5}. Therefore the remaining local calculation is the following
hypercube lemma.
\par
We first define the hypercube $HC$.
A $d$-dimensional hypercube $HC$ with its origin at $\vec{\square}$ is defined as a subspace with $2^d$  elements in $\mathbb{Z}_{\geq0}^n$:
\begin{equation}
\label{def:HC}HC^{(d)}(\vec{\square},\{\vec e_{n_i}\}_{i=1}^d)\coloneqq\{\tilde{\vec{\square}}\,|\,\tilde{\vec{\square}}=\vec{\square}+\sum_{i=1}^d \delta_i \vec{e}_{n_i},\delta_i=0,1\}.
\end{equation}

For this chosen set of directions, we denote the vertex opposite to the origin
by
\begin{equation}
\vec q_d\coloneqq\sum_{i=1}^d\vec e_{n_i}.
\end{equation}

\begin{figure}
    \centering
    \includegraphics[width=1\linewidth]{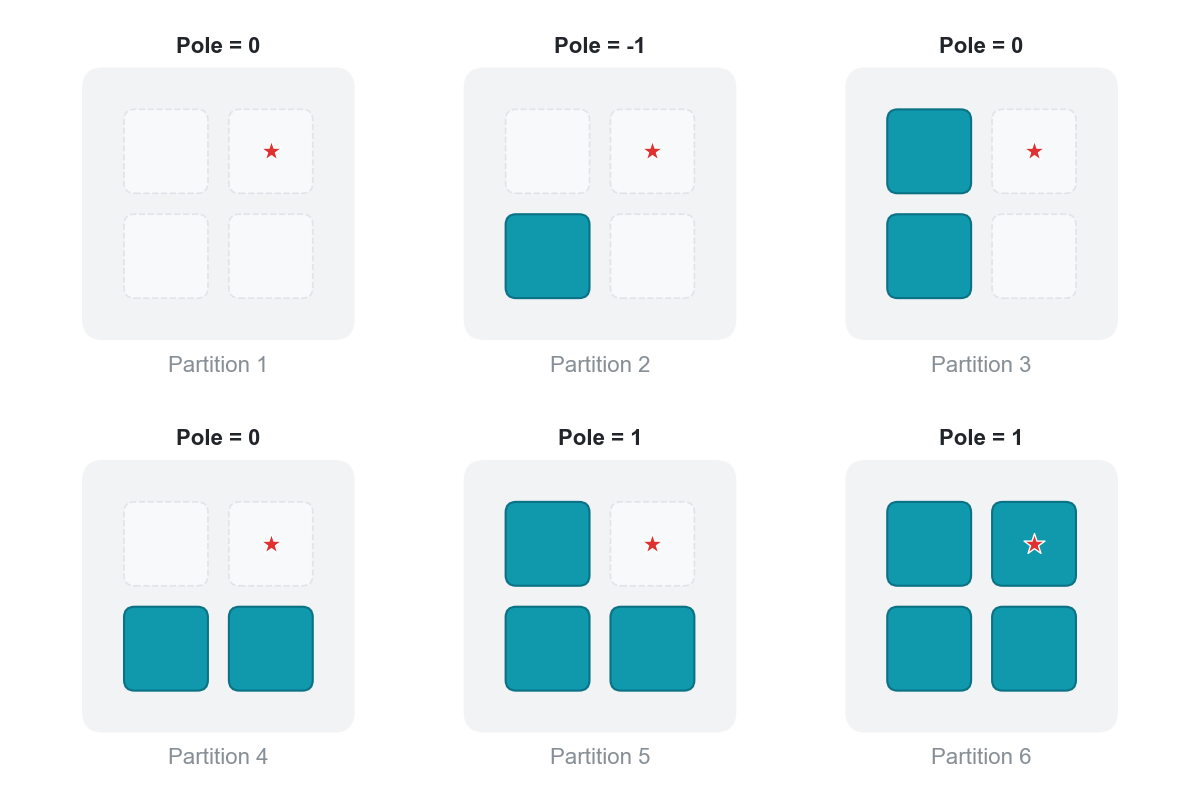}
    \caption{All six unique partitions of the hypercube ${HC}^{(2)}(n=6)$. Red star marks the target box position. Cyan color represents an occupied box.}
    \label{fig:HC2}
\end{figure}

For example, Fig.\ref{fig:HC2} shows all six unique partitions of the hypercube ${HC}^{(2)}(n=6)$.

\begin{Lemma}
(proved for 6D and discussed for higher n)
For every hypercube partition
$\Delta^{(n)}\in P_n$ with
$\Delta^{(n)}\subseteq HC^{(d)}(\vec{0},\{\vec{e}_{n_i}\})$,
\begin{empheq}[
  left={\begin{aligned}[b]
\omega_{0,\Delta^{(n)}}&(\vec q_d) \\
    &= \empheqlbrace
  \end{aligned}}
]{alignat=2}
    & \vphantom{\omega_{0,\Delta^{(n)}}}1,  &\quad& \text{if } \Delta^{(n)} \in G(\vec q_d), \tag{A1} \label{A1} \\
    & \le 0,  &\quad& \text{if } \Delta^{(n)} \notin G(\vec q_d). \tag{A2} \label{A2}
\end{empheq}
\label{Lemma5}
\end{Lemma}

To study Lemma \ref{Lemma5}, We write $HC^{(d)}\left(\vec{0},\{\vec{e}_{n_i}\}_{i=1}^d \right )=HC^{(d)}$ for short. We define the following set of admissible hypercube partitions:
\begin{equation}
\mathcal A_d
\coloneqq
\begin{cases}
\begin{gathered}
\bigl\{\emptyset,\{\vec 0\},\\
HC^{(n)}\setminus\{\vec q_n\},HC^{(n)}\bigr\},
\end{gathered}
& d=n,\\
\bigl\{HC^{(d)}\setminus\{\vec q_d\},HC^{(d)}\bigr\},
& d<n,
\end{cases}
\end{equation}
\begin{equation}
\Delta^{(n)}\in G(\vec q_d)
\iff \Delta^{(n)}\in\mathcal A_d.
\end{equation}

This characterization follows directly from the melting rule.

For ease of understanding, we can take Fig.\ref{fig:HC2} as an example. A simple pole occurs at the target position $\vec q_d$ only when either the target position alone remains unfilled or all positions are filled (Partition 5 and 6); otherwise, the potential is at most zero.

\subsection{\texorpdfstring{Analytic check of Lemma~\ref{Lemma5}}{Analytic check of Lemma A}}
We offer a proof for statement \eqref{A1}:
\begin{equation}
\Delta^{(n)}\in G(\vec q_d) \Rightarrow \omega_{0,\Delta^{(n)}}(\vec q_d)=1.
\end{equation}
First, it is direct to show:
\begin{equation}
\omega_{0,\emptyset}(\vec q_n)
=\omega_{0,\{\vec0\}}(\vec q_n)=1.
\end{equation}
\begin{equation}
\omega_{0,HC^{(d)}\setminus\{\vec q_d\}}(\vec q_d)
=\omega_{0,HC^{(d)}}(\vec q_d).
\end{equation}
To complete the proof of Lemma \ref{Lemma5}, it remains to prove:
\begin{equation}
\omega_{0,HC^{(d)}}(\vec q_d)=1.
\end{equation}

We introduce the concept of $d$-neighbor as follows. A projected point $x$ is called a $d$-neighbor of $x'$ (where $d<n$),
denoted by $x'\xrightarrow{d}x$, if and only if
\begin{equation}
x=x'+\sum_{i=1}^d h_{n_i},
\end{equation}
for some index set $\{n_i\}$ with $n_i \in \{1,2,\dots,n\}$.

Let $x_d\coloneqq c(\vec q_d)=\sum_{i=1}^d h_{n_i}$. The number of projected
points in the hypercube satisfying the $m$-neighbor relation with $x_d$ is
\begin{equation}
\left|
\left\{x=c(\vec{\square})\ \middle|\
\vec{\square}\in HC^{(d)},\ x\xrightarrow{m}x_d
\right\}
\right|
=\binom{d}{m}.
\end{equation}

For each $k\in\{3,5,\ldots,2K-3\}$, a cluster centered at an
$(n-k)$-neighbor contributes a pole at $x_d$. Each $(n-1)$-neighbor
contributes one zero through its single-box factor and two poles through its
cluster factor, giving one net pole. Altogether, an $m$-neighbor contributes
$(-1)^{m+1}$ to the potential at $x_d$.

Case 1: $d<n$
\begin{equation}
\omega_{HC^{(d)}}(\vec q_d)
=\sum_{m=1}^d\binom{d}{m}(-1)^{m+1}=1,
\end{equation}
and
\begin{equation}
\omega_{0,HC^{(d)}}(\vec q_d)=\omega_{HC^{(d)}}(\vec q_d).
\end{equation}

Case 2: $n = d$
\begin{equation}
\begin{aligned}
\omega_{HC^{(n)}}(\vec q_n)
&=\sum_{m=1}^{n-1}\binom{n}{m}(-1)^{m+1}-2\\
&=\sum_{m=0}^{n}\binom{n}{m}(-1)^{m+1}=0.
\end{aligned}
\end{equation}
\begin{equation}
\omega_{0,HC^{(n)}}(\vec q_n)
=1+\omega_{HC^{(n)}}(\vec q_n)=1.
\end{equation}
Thus we have completed the proof of statement \eqref{A1} in Lemma \ref{Lemma5}.

\subsection{Numerical Verification}
\label{sec:numerical}

For the statement \eqref{A2}, we do not provide an analytical proof. Instead, we verify it for 6D by exhaustive enumeration. For 8D, we perform Monte Carlo sampling tests to check its correctness.
\begin{figure*}[t]
    \centering

    \begin{minipage}{0.49\textwidth}
        \centering
        \includegraphics[width=\linewidth]{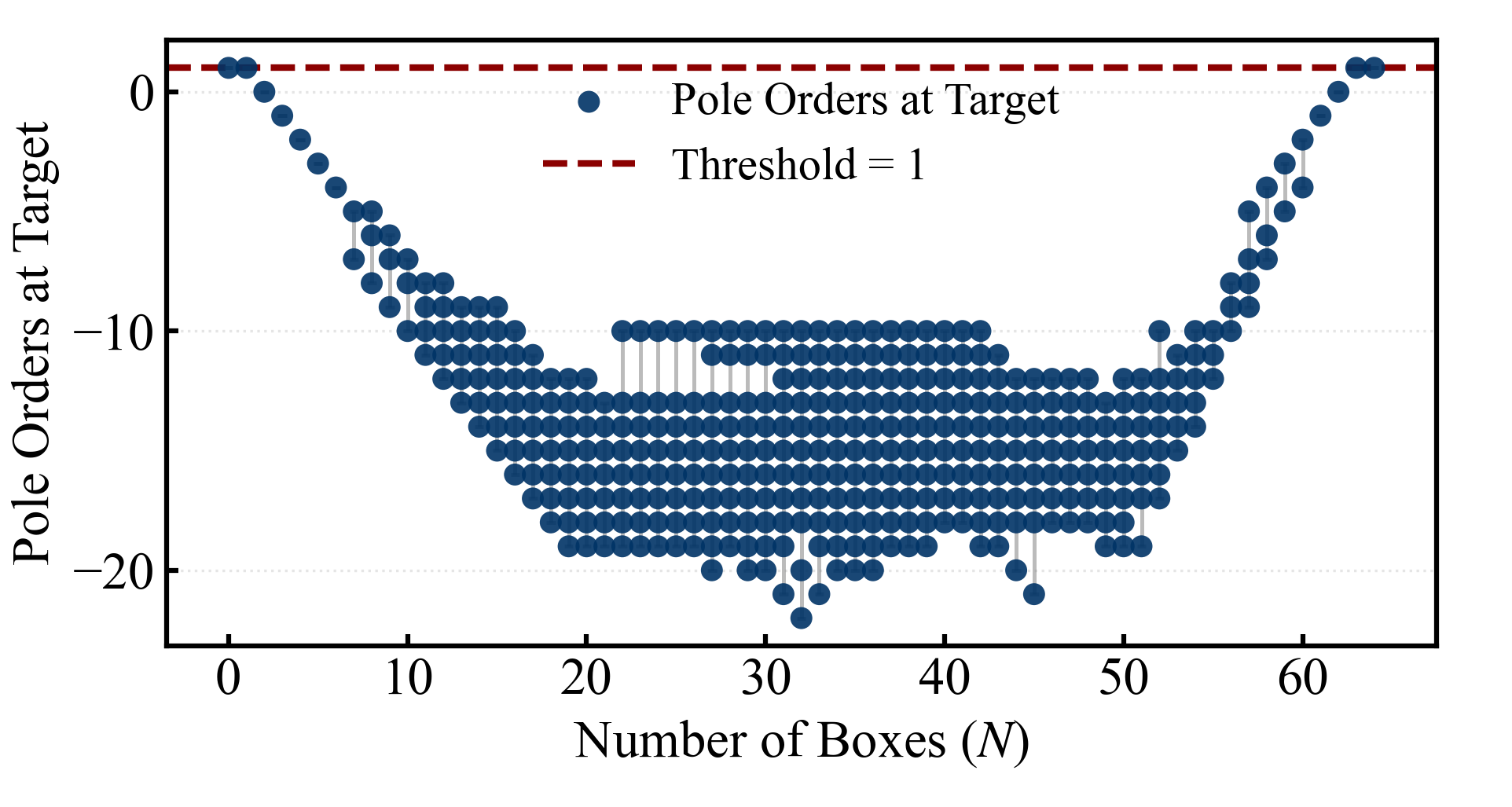}
        \par\vspace{3pt}
        (a) Exhaustive enumeration for $n=6$.
    \end{minipage}
    \hfill
    \begin{minipage}{0.49\textwidth}
        \centering
        \includegraphics[width=\linewidth]{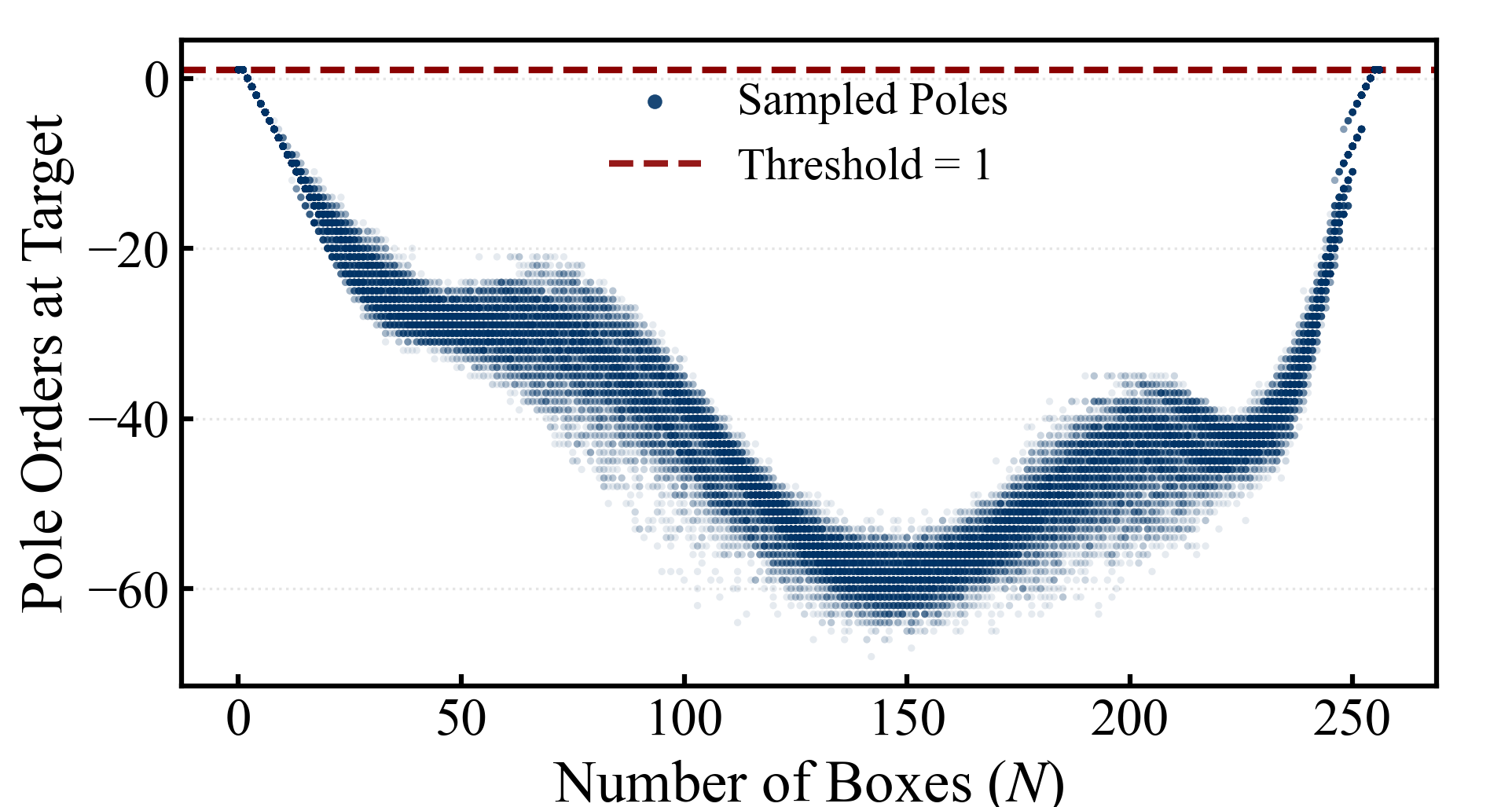}
        \par\vspace{3pt}
        (b) Monte Carlo sampling for $n=8$.
    \end{minipage}
    \caption{Numerical verification of the pole order at the target position $\vec{\square}=(1,1,\dots,1)$ for even dimensional partitions within a hypercube. The red dashed line represents the theoretical threshold $\omega_{0,\Delta^{(n)}}=1$. (a) The 6D case shows all possible partitions, forming a clear boundary. (b) The 8D case, obtained via sampling, shows a consistent distribution where all sampled states satisfy $\omega_{0,\Delta^{(n)}}\leq 1$. Note that the pole order equals 1 only for the limiting cases ($N=0, 1, 2^n-1, 2^n$).}
    \label{fig:even_dim_check}
\end{figure*}

For the 6D case, the total number of boxes in the hypercube $HC^{(6)}$ is $2^6 = 64$. The size of the configuration space is manageable (in total 7836132 cases), allowing us to perform an exhaustive enumeration of all possible partitions $\Delta^{(6)} \subseteq HC^{(d)},(d=1,2,...,6)$. We calculate the pole order $\omega_{0,\Delta^{(6)}}(\vec{\square})$ at the target position $\vec{\square}=(1,1,\dots,1)$ for every unique partition.

The results for $d=6$ are presented in Fig.\ref{fig:even_dim_check}(a). The horizontal axis represents the number of boxes $N = |\Delta^{(6)}|$, ranging from 0 to 64. The vertical axis shows the calculated pole order. It is evident that for all valid partitions, the pole order satisfies the condition $\omega_{0,\Delta^{(6)}}\leq 1$. The maximum value $\omega_{0,\Delta^{(6)}}=1$ is achieved only when $N=0,1,63,64$ , which is consistent with the theoretical prediction that these are the only admissible configurations in $G(\vec q_n)$ for the hypercube geometry. The distribution of pole orders forms a characteristic ``U'' shape, strictly bounded by the threshold line $\omega_{0,\Delta^{(6)}}=1$.

For the 8D case, the hypercube $HC^{(8)}$ contains $2^8 = 256$ boxes. The number of unique partitions (equal to the Dedekind number $M(8)$) is astronomically large, making exhaustive enumeration computationally intractable. Therefore, we adopted an independent Monte Carlo sampling approach. For each box number $N \in [0, 256]$, we generated a large number of random valid partitions satisfying the melting rule and computed their potential functions.

Figure~\ref{fig:even_dim_check}(b) displays the results of the 8D sampling. Similar to the 6D case, the sampled pole orders are densely distributed below the threshold line. Despite the significant increase in dimensionality and complexity, no violations of the upper bound $\omega_{0,\Delta^{(8)}}\le 1$ were observed. The density of the scatter plot (represented by the transparency in the figure) indicates the statistical distribution of the pole orders, confirming that the vast majority of configurations lie well within the negative region, while the boundary conditions at $N=0,1,255,256$ correctly return to $\omega_{0,\Delta^{(8)}}=1$.

These numerical results provide strong evidence supporting our formula for even dimensional partitions, confirming that the constructed charge function correctly reproduces the pole structure required by the BPS algebra representation theory.


\section{Conclusion}
By combining the even-dimensional charge functions ($n=2K, K \ge 2$) proposed in this letter with our previously established results for odd dimensions, together with the lower dimensional cases in the literature, we have completed the library of charge functions for partitions in all dimensions, providing the essential foundation for exploring the BPS algebras of higher-dimensional toric Calabi-Yau manifolds.

\begin{acknowledgments}
The authors thank Yutaka Matsuo, Jean-Emile  Bourgine and Rui-Dong Zhu
for helpful discussions.  K.Z. (Hong Zhang) is supported by a classified fund of Shanghai city.
\end{acknowledgments}

\appendix

\section{\texorpdfstring{Reduction to Hypercube Geometry}{Reduction to Hypercube Geometry}}
\label{appendix:hypercube_reduction}
We summarize the reduction step used before Lemma~\ref{Lemma5}. The purpose
is not to give a second complete proof of all auxiliary statements, but to
make clear how the hypercube lemma enters the induction. The detailed
combinatorial proofs of the auxiliary bisect statements are the same as those
given for Lemmas 1--4 in Ref.~\cite{xiang2025chargefunctionshighdimensional}.

Fix a partition $\Delta^{(n)}$. To relate the melting rule in the lattice
$\mathbb Z_{\geq0}^{n}$ to the pole structure of the charge function in the
projected space $\mathcal P$, we use the surface set $F$. The word
``surface'' means that, for a given projected point, we choose the
non-occupied lattice representative closest to the origin along the diagonal
direction. Equivalently, this representative cannot be shifted backward by
$\vec q_n$ while remaining outside $\Delta^{(n)}$. Such a point represents a
candidate addable/removable projected position, but it is not necessarily
itself a legal addable or removable box.

We now recall the precise surface set notation used in
Ref.~\cite{xiang2025chargefunctionshighdimensional}. For a fixed
$d\leq n$ and a set of coordinate directions
$\{\vec e_{n_i}\}_{i=1}^{d}$, define
\begin{equation}
\label{eq:F_surface_appendix}
F(d,\{\vec e_{n_i}\})
\coloneqq
\begin{cases}
F_{<n}(d,\{\vec e_{n_i}\}), & \text{if } d<n,\\
F_n, & \text{if } d=n,
\end{cases}
\end{equation}
Here $F_{<n}(d,\{\vec e_{n_i}\})$ is the set of all
$\vec{\square}=\sum_i l_i\vec e_i$ such that
$\vec{\square}\notin\Delta^{(n)}$, $l_{n_j}>0$ for
$j=1,\ldots,d$, and $l_k=0$ for
$k\notin\{n_1,\ldots,n_d\}$. The set $F_n$ is the set of all
$\vec{\square}=\sum_i l_i\vec e_i$ such that
$\vec{\square}\notin\Delta^{(n)}$ and
$\vec{\square}-\vec q_n\in\Delta^{(n)}$.
For $d<n$, the integer $d$ is the number of nonzero coordinate directions of
the surface position $\vec{\square}$ itself: the nonzero coordinates are
precisely those in $\{n_1,\ldots,n_d\}$, and the other $n-d$ coordinates are
zero. The case $d=n$ is the diagonal surface case, where all $n$ directions
enter through the condition $\vec{\square}-\vec q_n\in\Delta^{(n)}$.

For a point $\vec a=\sum_i a_i\vec e_i$, define the bisect operation by
\begin{equation}
\begin{aligned}
L(\Delta^{(n)},\vec a)
&\coloneqq
\{\vec b=\sum_i b_i\vec e_i\in\Delta^{(n)}\mid \\
&\hspace{3.5em}
b_i\geq a_i,\quad i=1,\ldots,n\}.
\end{aligned}
\end{equation}
For a surface position
$\vec{\square}\in F(d,\{\vec e_{n_i}\})$, write
$\vec q_d=\sum_{i=1}^{d}\vec e_{n_i}$. We then consider the hypercube
associated with this surface position, using the definition
\eqref{def:HC} with origin $\vec{\square}-\vec q_d$ and directions
$\{\vec e_{n_i}\}_{i=1}^{d}$. Set
\begin{equation}
\begin{aligned}
\widetilde L
&=L(\Delta^{(n)},\vec{\square}-\vec q_d),\\
\widetilde{HC}
&=HC^{(d)}(\vec{\square}-\vec q_d,\{\vec e_{n_i}\}_{i=1}^{d})
\cap \Delta^{(n)} .
\end{aligned}
\end{equation}
Here $\widetilde{HC}$ is the part of the partition contained in this
elementary hypercube. The larger set $\widetilde L$ is the finite upper part
which must be removed if one wants the remaining set
$\Delta^{(n)}-\widetilde L$ to remain a partition. In other words, we are not
removing only the hypercube boxes; we also remove the boxes whose presence
would violate the melting rule after the hypercube part is removed.

The following technical facts are used in the induction. First,
$\Delta^{(n)}-\widetilde L$ is again a partition. Second, the boxes in
$\widetilde L-\widetilde{HC}$ do not change the pole order at
$c(\vec{\square})$, so
\begin{equation}
\omega_{\Delta^{(n)}}(\vec{\square})
=
\omega_{\Delta^{(n)}-\widetilde L+\widetilde{HC}}(\vec{\square}) .
\end{equation}
Third, the cluster terms involving both $\widetilde{HC}$ and
$\Delta^{(n)}-\widetilde L$ have zero contribution at $c(\vec{\square})$.
Thus the pole order decomposes into a contribution from the smaller partition
and a contribution from the elementary hypercube:
\begin{equation}
\omega_{\Delta^{(n)}-\widetilde L+\widetilde{HC}}(\vec{\square})
=
\omega_{\Delta^{(n)}-\widetilde L}(\vec{\square})
+\omega_{\widetilde{HC}}(\vec{\square}) .
\end{equation}

These facts are precisely where the technical details of the bisect argument
enter. They depend only on the melting rule and on the projected neighborhood
relation; the full verification is given in Ref.~\cite{xiang2025chargefunctionshighdimensional}.

We now explain how this gives the induction step. Since
$|\Delta^{(n)}-\widetilde L|<|\Delta^{(n)}|$, the smaller partition can be
handled by the induction hypothesis whenever its contribution is nontrivial.
When $d<n$, the background contribution
$\omega_{0,\Delta^{(n)}-\widetilde L}(\vec{\square})$ vanishes, so this part
is already trivial. When $d=n$, the Calabi--Yau relation identifies
$c(\vec{\square})$ with $c(\vec{\square}-\vec q_n)$, and the induction
hypothesis applies to the smaller partition at the corresponding boundary
position. Therefore the only contribution not fixed by induction is
$\omega_{\widetilde{HC}}(\vec{\square})$.

Finally, by translating the elementary hypercube so that
$\vec{\square}-\vec q_d$ becomes the origin, the remaining local calculation
has the standard form
\begin{equation}
\Delta^{(n)}\subseteq HC^{(d)}(\vec 0,\{\vec e_{n_i}\}_{i=1}^{d}),
\qquad
\text{evaluated at } \vec q_d .
\end{equation}
This is the situation covered by Lemma~\ref{Lemma5}. Thus the role of the
hypercube is not that every partition is repeatedly reduced until only a
hypercube remains. Rather, in one induction step we cut from the original
partition a finite upper part determined by an elementary hypercube; the
remaining partition is either handled by induction or gives a trivial
background contribution, and Lemma~\ref{Lemma5} evaluates the finite local
difference.

\section{Lower Dimensions}
\label{appendix:lower_dim}

We recall the known lower-dimensional charge functions in the notation of the
present paper, following the standard formulas given in
Ref.~\cite{galakhov2024charging}.

\paragraph{Case $n=2$: Young Diagrams}
For 2D partitions, the Calabi-Yau condition specializes to
\begin{equation}
    h_1+h_2=0.
\end{equation}
The corresponding charge function associated with the Young diagram $\Delta^{(2)}$ is given by:
\begin{equation}
\begin{split}
\label{2D charge function}
\psi_{\Delta^{(2)}}(u) = \frac{1}{u}
& \prod_{\vec{\square} \in \Delta^{(2)}} \varphi_1^{(2)} \left( u - c(\vec{\square}) \right) \\
\times \prod_{\phi_{2} \subseteq \Delta^{(2)}} &\varphi_{2}^{(2)} \left( u - b(\phi_{2}) \right)
\prod_{\phi_3 \subseteq \Delta^{(2)}} \varphi_3^{(2)} \left( u - c(\phi_3) \right),
\end{split}
\end{equation}
where the factors are defined as
\begin{equation}
\begin{split}
   &\varphi_{1}^{(2)}(u)=\frac{1}{(u-h_1)(u-h_2)},\\
   &\varphi_2^{(2)}(u)=u^2,\quad \varphi_3^{(2)}(u)=\frac{1}{u^2}.
\end{split}
\end{equation}

Here the two-dimensional doublet correction uses the standard base-centered
coordinate. More precisely, for
$\phi_2(\vec{\square};i)=\{\vec{\square},\vec{\square}+\vec e_i\}$
with $i=1,2$, we define
\begin{equation}
    b(\phi_2(\vec{\square};i))
    \coloneqq c(\vec{\square}).
\end{equation}

This two-dimensional convention is special to the doublet factor and should
not be confused with the shifted cluster coordinate $c(\phi_p)$ introduced
above.

\vspace{1em}
\paragraph{Case $n=3$: Plane Partitions}
For 3D (plane) partitions, the Calabi-Yau constraint takes the form
\begin{equation}
    h_1+h_2+h_3=0.
\end{equation}
The charge function for $\Delta^{(3)}$ is expressed in terms of a single species of factor $\varphi_1^{(3)}(u)$:
\begin{equation}
\psi_{\Delta^{(3)}}(u) = \frac{1}{u} \prod_{\vec{\square} \in \Delta^{(3)}} \varphi^{(3)}_1(u - c(\vec{\square})), \label{3D charge function}
\end{equation}
where
\begin{equation}
\varphi^{(3)}_1(u) = \prod_{i=1}^{3} \frac{u + h_i}{u - h_i}.
\end{equation}
This 3D structure is known to be governed by the affine Yangian of $\mathfrak{gl}_{1}$ .

\vspace{1em}
\paragraph{Case $n=4$: Solid Partitions}
In the 4D case, the Calabi-Yau condition reads
\begin{equation}
    h_1+h_2+h_3+h_4=0.
\end{equation}
The charge function for the solid partition $\Delta^{(4)}$ incorporates both single-box and cluster contributions:
\begin{equation}
\begin{split}
\psi_{\Delta^{(4)}}(u) = \frac{1}{u}
& \prod_{\vec{\square} \in \Delta^{(4)}} \varphi_1^{(4)} \left( u - c(\vec{\square}) \right) \\
\times \prod_{\phi_{4} \subseteq \Delta^{(4)}} &\varphi_{4}^{(4)} \left( u - c(\phi_{4}) \right)
\prod_{\phi_5 \subseteq \Delta^{(4)}} \varphi_5^{(4)} \left( u - c(\phi_5) \right),
\end{split}
\label{4D charge function}
\end{equation}
where the factors are defined by
\begin{equation}
\begin{aligned}
   \varphi_{1}^{(4)}(u) &= \frac{\prod_{i=1}^{4}(u+h_i)\prod_{1\le i<j\le 4}(u-h_i-h_j)}{\prod_{i=1}^{4}(u-h_i)}, \\
   \varphi_4^{(4)}(u) &= \frac{1}{u^2}, \quad \varphi_5^{(4)}(u) = u^2.
\end{aligned}
\end{equation}
\vspace{1em}
The factor $\varphi_{1}^{(4)}(u)$ also admits the following symmetric representation:
\begin{equation}
\begin{aligned}
   \varphi_{1}^{(4)}(u) = \frac{\prod\limits_{1\le i<j\le 4}(u-h_i-h_j)}{\prod\limits_{i=1}^{4}(u-h_i)} \\\times \prod\limits_{1\le i<j<k\le 4}(u-h_i-h_j-h_k).
\end{aligned}
\end{equation}

\section{\texorpdfstring{Simple Low-Dimensional Examples}{Simple Low-Dimensional Examples}}
\label{appendix:low_dim_examples}
We give a few elementary examples to illustrate how the poles of the charge
function reproduce the projected addable and removable box positions. In each
dimension we consider the empty partition
$\Delta_0=\emptyset$, the one-box partition
$\Delta_1=\{\vec0\}$, and the two-box partition
$\Delta_2=\{\vec0,\vec e_1\}$. The other one-direction two-box examples are
obtained by permuting the indices.

\paragraph{Case $n=2$.}
Here $h_1+h_2=0$. For the empty Young diagram,
\begin{equation}
    \psi_{\Delta_0}^{(2)}(u)=\frac{1}{u},
\end{equation}
which has a pole at $u=0$, corresponding to the addable box $\vec0$.
For the one-box diagram,
\begin{equation}
    \psi_{\Delta_1}^{(2)}(u)
    =\frac{1}{u}\varphi_1^{(2)}(u)
    =\frac{1}{u(u-h_1)(u-h_2)}.
\end{equation}
Thus the poles are at $0,h_1,h_2$, corresponding respectively to the
removable box $\vec0$ and the addable boxes $\vec e_1,\vec e_2$.
For $\Delta_2=\{\vec0,\vec e_1\}$, the standard two-dimensional doublet
factor is centered at the base box of the pair, namely
$b(\phi_2(\vec0;1))=c(\vec0)=0$. Hence
\begin{equation}
\begin{aligned}
    \psi_{\Delta_2}^{(2)}(u)
    &=\frac{1}{u}\varphi_1^{(2)}(u)
      \varphi_1^{(2)}(u-h_1)\,\varphi_2^{(2)}(u)  \\
    &=\frac{1}{(u-h_1)(u-h_2)(u-2h_1)} .
\end{aligned}
\end{equation}
The poles are therefore at $h_1,h_2,2h_1$, which are the projected positions
of the removable box $\vec e_1$ and the addable boxes $\vec e_2,2\vec e_1$.

\paragraph{Case $n=3$.}
Here $h_1+h_2+h_3=0$ and
\[
    \varphi_1^{(3)}(u)=\prod_{i=1}^{3}\frac{u+h_i}{u-h_i}.
\]
For the empty partition,
\begin{equation}
    \psi_{\Delta_0}^{(3)}(u)=\frac{1}{u},
\end{equation}
so the only pole corresponds to adding $\vec0$. For the one-box partition,
\begin{equation}
    \psi_{\Delta_1}^{(3)}(u)
    =\frac{1}{u}\varphi_1^{(3)}(u),
\end{equation}
whose poles are at $0,h_1,h_2,h_3$. These are the projected positions of the
removable box $\vec0$ and the addable boxes $\vec e_1,\vec e_2,\vec e_3$.
For $\Delta_2=\{\vec0,\vec e_1\}$,
\begin{equation}
\begin{aligned}
    \psi_{\Delta_2}^{(3)}(u)
    &=\frac{1}{u}\varphi_1^{(3)}(u)\varphi_1^{(3)}(u-h_1)\\
    &=\frac{(u+h_1)(u-h_1+h_2)(u-h_1+h_3)}
    {(u-h_1)(u-h_2)(u-h_3)(u-2h_1)} .
\end{aligned}
\end{equation}
Thus the pole positions are $h_1,h_2,h_3,2h_1$, corresponding to the
removable box $\vec e_1$ and the addable boxes
$\vec e_2,\vec e_3,2\vec e_1$.

\paragraph{Case $n=4$.}
Let
\[
    \Phi_1^{(4)}(u)
    =
    \frac{\prod_{i=1}^{4}(u+h_i)\prod_{1\le i<j\le 4}(u-h_i-h_j)}
    {\prod_{i=1}^{4}(u-h_i)} ,
\]
with $h_1+h_2+h_3+h_4=0$. For the empty solid partition,
\begin{equation}
    \psi_{\Delta_0}^{(4)}(u)=\frac{1}{u},
\end{equation}
whose pole corresponds to adding $\vec0$. For the one-box partition,
\begin{equation}
    \psi_{\Delta_1}^{(4)}(u)
    =\frac{1}{u}\Phi_1^{(4)}(u),
\end{equation}
and the poles are at $0,h_1,h_2,h_3,h_4$, corresponding to the removable box
$\vec0$ and the addable boxes $\vec e_1,\vec e_2,\vec e_3,\vec e_4$.
For $\Delta_2=\{\vec0,\vec e_1\}$, no four-box or five-box cluster is present,
so
\begin{equation}
    \psi_{\Delta_2}^{(4)}(u)
    =
    \frac{1}{u}\Phi_1^{(4)}(u)\Phi_1^{(4)}(u-h_1).
\end{equation}
Using the Calabi--Yau relation, the numerator cancels the possible spurious
poles at $u=0$ and $u=h_1+h_j$ for $j=2,3,4$. The remaining poles are
\begin{equation}
    u=h_1,\ h_2,\ h_3,\ h_4,\ 2h_1,
\end{equation}
which are precisely the projected positions of the removable box $\vec e_1$
and the addable boxes $\vec e_2,\vec e_3,\vec e_4,2\vec e_1$.

\bibliography{FCZ}

\end{document}